\documentclass[twoside,british,3p]{elsarticle}
\usepackage[T1]{fontenc}
\usepackage[latin9]{inputenc}
\usepackage[table,xcdraw]{xcolor}
\usepackage{pagecolor}
\usepackage{pdfcolmk}
\usepackage{units}
\usepackage{textcomp}
\usepackage{bbding}
\usepackage{amsmath}
\usepackage{amssymb}
\usepackage{graphicx}
\usepackage{setspace}
\usepackage{esvect}
\PassOptionsToPackage{normalem}{ulem}
\usepackage{ulem}
\usepackage{blindtext}
\usepackage{multicol}
\usepackage{booktabs}
\usepackage{tabularx}
\usepackage{longtable}
\usepackage{pbox}
\usepackage{csquotes}

\usepackage{caption}
\usepackage{subcaption}

\usepackage{lscape}
\usepackage{hyperref}
\usepackage{makecell}
\usepackage{multirow}

\usepackage{enumitem}

\usepackage{natbib}

\newcount\savedtheorem

\hypersetup{
    colorlinks=true,
    linkcolor=blue,
    filecolor=magenta,      
    urlcolor=cyan,
    pdfborderstyle={/S/U}
}

\newcommand{\Tau}{\mathcal{T}}

\makeatletter

\providecommand{\tabularnewline}{\\}

\journal{JORS}
\bibpunct{(}{)}{;}{a}{,}{,} 

\DeclareGraphicsExtensions{.pdf,.jpeg,.png,.eps}

\usepackage{dcolumn}   
\usepackage{bm}        

\usepackage{babel}

\begin{document}

\begin{frontmatter}{}

\title{Preference Analysis Using Random Spanning Trees: A Stochastic Sampling Approach to Inconsistent Pairwise Comparisons}

\author[aff1]{Salvatore Greco}
\ead{salgreco@unict.it}
\author[aff2]{Sajid Siraj\corref{fn1}}
\ead{s.siraj@leeds.ac.uk}
\author[aff3]{Michele Lundy}
\ead{michele.lundy@plymouth.ac.uk}

\cortext[fn1]{Room 1.22, Charles Thackrah Building, Clarendon Road, Leeds University Business School, Leeds LS2 9JT, UK}

\address[aff1]{Department of Economics, University of Catania, Catania, Sicily, Italy}
\address[aff2]{Leeds University Business School, University of Leeds, Leeds, UK}
\address[aff3]{University of Plymouth, Plymouth, UK}

\begin{abstract}
Eliciting preferences from human judgements is inherently imprecise, yet most decision analysis methods force a single priority vector from pairwise comparisons, discarding the information embedded in inconsistencies. We instead leverage inconsistency to characterise preference uncertainty by examining all priority vectors consistent with the decision maker's judgements. Spanning tree analysis enumerates combinations of evaluation and weighting vectors from pairwise comparison subsets, each yielding a distinct preference vector and collectively defining a distribution over possible preference orderings. Since exponential growth renders complete enumeration prohibitive, we propose a stochastic random walk sampling approach with sample sizes formally established via statistical sampling theory. This enables two key metrics: Pairwise Winning Indices (PWIs), the probability one alternative is preferred to another, and Rank Acceptability Indices (RAIs), the probability an alternative attains a given rank. A notable advantage is applicability to incomplete pairwise comparisons, common in large-scale problems. We validate the methodology against complete enumeration on a didactic example, then demonstrate scalability on a telecommunications backbone infrastructure selection case study involving billions of spanning tree combinations. The approach yields probabilistic insights into preference robustness and ranking uncertainty, supporting informed decisions without the burden of exact enumeration.\\
\textbf{Practitioner Summary}\\
When making complex decisions - such as choosing between infrastructure options or selecting a school - decision-makers are often asked to compare alternatives in pairs and rate their relative importance. This structured approach, known as pairwise comparison, is widely used in business, government, and policy settings. However, human judgements are rarely perfectly consistent: a decision-maker might rate Option A twice as good as B, and B three times as good as C, but then rate A only four times as good as C rather than the mathematically expected six. Rather than treating such inconsistencies as errors to be corrected, this paper treats them as meaningful signals that the decision-maker holds multiple valid perspectives simultaneously.
We develop a practical method that captures this uncertainty and translates it into useful probability-based outputs. Specifically, for any pair of options, our method estimates the probability that one is genuinely preferred over the other. It also estimates the probability that each option would rank first, second, third, and so on. These outputs help practitioners understand not just which option "wins," but how confident they can be in that conclusion.
A key practical advantage is that the method works even when some comparisons are missing - common in large real-world problems where decision-makers cannot respond to every possible pair. We demonstrate the approach on a telecommunications infrastructure selection problem involving billions of possible combinations, showing that reliable probability estimates can be obtained from a manageable sample of around 20,000 iterations.
\end{abstract}

\begin{keyword}
Pairwise comparisons; Spanning trees; Random sampling; Preference uncertainty; Stochastic acceptability analysis

\end{keyword}

\end{frontmatter}

Word Count: 7,529

\section{Introduction\label{sec:intro}}

Decision problems invariably require systematic comparison of multiple alternatives across several criteria. Consider, for example, vehicle selection: one must evaluate competing models with respect to salient attributes such as price, performance, and aesthetics. A rigorous approach to such decisions typically involves three sequential stages:
\begin{enumerate}
\item[(1)] assign relative priorities to each alternative with respect to individual criteria, establishing performance profiles;
\item[(2)] establish the relative importance of criteria to reflect decision-maker preferences;
\item[(3)] aggregate alternative evaluations across criteria using the derived weights to produce a comprehensive preference assessment.
\end{enumerate}

Pairwise comparison represents a well-established methodological approach for assigning such priorities. Its principal appeal lies in cognitive efficiency: by requiring decision-makers to focus on two items sequentially rather than the complete set simultaneously, it mitigates information processing constraints inherent to human judgment.

\subsection{The Consistency Challenge}

When three or more items require prioritisation, direct comparison between all pairs frequently yields judgments differing from mathematically derived results. This divergence---termed \textit{inconsistency}---arises from diverse sources: psychological limitations such as bounded cognition, procedural constraints such as restricted response scales (e.g., 1--9 intervals), or legitimate context-dependent changes in preference.

The literature distinguishes two consistency requirements. \textit{Ordinal consistency} requires that preference orderings exhibit transitivity (e.g., if Price is more important than Speed, and Speed more important than Looks, then Price must be more important than Looks). \textit{Cardinal consistency} imposes a stricter requirement: if Price is preferred to Speed $p$ times, and Speed to Looks $q$ times, then Price should be preferred to Looks $pq$ times. For example, if Price is twice as important as Speed and Speed is three times as important as Looks, cardinal consistency requires Price to be six times as important as Looks. When this condition is violated (e.g., the decision-maker assigns three times instead), judgements are cardinally inconsistent, though ordinal consistency may still be preserved. Whilst cardinally consistent judgments guarantee ordinal consistency, the converse does not obtain.

\subsection{Extant Methods and Their Limitations}

Numerous techniques have been proposed for eliciting priority vectors from inconsistent judgments---including the row geometric mean \citep{Crawford1985}, logarithmic least squares \citep{deJong1984}, the eigenvector method \citep{Saaty1977}, and spanning tree aggregation approaches \citep{Siraj2012,Tsyganok2010}. Conventionally, these methods seek to discover underlying ``true preferences'' by amalgamating observed judgments, effectively treating inconsistency as measurement error to be minimised.

The Analytic Hierarchy Process (AHP), the most widely adopted pairwise comparison methodology, derives priorities via the Right Eigenvector (REV) method. Whilst computationally straightforward, REV has attracted sustained criticism: left-right eigenvector asymmetries, arbitrary inconsistency thresholds (e.g., the conventional CR $< 0.1$ criterion), and potential preference reversals in certain contexts \citep{Brunelli2018}. Although the Row Geometric Mean method is increasingly prevalent and mathematically tractable, it encounters analogous limitations. Comprehensive comparative analyses \citep{Choo2004,Brunelli2018} demonstrate conclusively that no universally superior method exists across all decision contexts.

\subsection{A Reconceptualisation: Preference Uncertainty Rather Than Error}

We propose a fundamentally distinct perspective: rather than treating inconsistency as error requiring minimisation, we interpret it as evidence of preference uncertainty---a manifestation of the inherent complexity and multifaceted nature of real-world preference structures. This reconceptualisation is grounded in robust findings from behavioural economics demonstrating that individual preferences frequently exhibit context-dependence and multiplicity \citep{Ainslie2001,Schelling1980}. When individuals express divergent judgments about the same preference relationship, this plurality need not reflect measurement error or cognitive failure; rather, it may reveal the coexistence of multiple coherent preference orderings within a single decision-maker.

The spanning trees approach offers a graph-theoretic mechanism for implementing this perspective. By systematically enumerating all possible combinations of independent judgments (spanning trees) within a pairwise comparison matrix, one obtains the complete set of feasible priority vectors, each representing a distinct coherent preference ordering. Hitherto, researchers have amalgamated these vectors into a single solution using arithmetic or geometric averaging, thereby discarding the information embodied in their distribution. We propose instead to preserve and analyse this plurality, computing probability distributions over preference orderings.

However, the exponential proliferation of spanning tree combinations renders complete enumeration computationally prohibitive for realistic problem sizes. A decision problem with merely four criteria and four alternatives yields over one million possible combinations of priority vectors. This computational barrier has prevented widespread adoption of the spanning trees approach despite its conceptual attractions.

\subsection{Contribution and Paper Structure}

To address this limitation, we propose a stochastic approach employing random walk sampling to generate statistically representative samples of spanning tree combinations without complete enumeration. Building upon recent developments in stochastic multi-criteria acceptability analysis (SMAA), we compute two key metrics: Pairwise Winning Indices (PWIs)---the probability that one alternative is preferred to another---and Rank Acceptability Indices (RAIs)---the probability distributions over ranking positions. Using standard statistical sampling theory, we establish how many iterations are required to achieve specified accuracy levels.

A notable advantage of this approach is its direct applicability to incomplete pairwise comparison matrices, a common occurrence when decision-makers cannot or will not provide responses to all possible comparisons. Existing methods typically require either imputation of missing values or complete data; our approach accommodates incomplete data without modification.

The paper is organised as follows. Section \ref{sec:background} provides background on pairwise comparisons and the spanning trees approach. Section \ref{sec:spanning} formalises the stochastic framework and introduces PWI and RAI metrics. Section \ref{sec:randomspanning} develops the random sampling procedure and establishes its statistical properties. Section \ref{sec:telecomstudy} demonstrates the methodology on a realistic infrastructure selection problem. Section \ref{sec:conclusion} concludes with discussion of limitations and directions for future research.

\section{Background\label{sec:background}}
Pairwise comparison is a commonly used approach to obtain a DM's priorities and ultimately overall preferences. For example, in a car selection problem, if we consider three criteria of Price, Speed, and Looks, we can either ask the DM to directly assign weights to these three criteria; or we can alternatively ask him/her to compare two criteria at a time and make these comparisons for all three possible pairs, that is, Price versus Speed, Speed versus Looks, and then Price versus Looks. These comparative judgements can then be used to calculate the relative weights in the form of a weight vector. 

Whilst comparing the three criteria, one would expect that if Price is considered more important than Speed and Speed more important than Looks, then Price must be more important than Looks. This is often referred to as the ordinal consistency requirement \citep{Kwiesielewicz2004}. A much stronger requirement would be that if Price is preferred to Speed $p$ times, and Speed is preferred to Looks $q$ times, then Price should be preferred to Looks $pq$ times. This is often termed cardinal consistency \citep{Saaty1980, siraj2015contribution}. Although these are rational and justified requirements from a mathematical perspective, DMs often break these rules. Obviously, if the DM is cardinally consistent, then the ordinal consistency will already be achieved. However, the opposite is not true, that is, ordinal consistency does not ensure that the judgements are cardinally consistent.

This is demonstrated through the car example where if Price is considered twice as important as Speed, and Speed is considered three times as important as Looks then one might suggest that  Price is 6 times as important as Looks. If this is indeed the case then we can easily calculate the weight vector, however, if instead of choosing 6, the DM suggests Price is 3 times as important as Looks, then this makes the three judgements cardinally inconsistent - though note that the order of preference is still preserved and so the judgements are not ordinally inconsistent. There can be many different reasons for having inconsistent judgements, including but not limited to psychological or procedural limitations. An example of a psychological reason would be the limited capacity of the human brain to process information for comparing items \citep{saaty2003magic}. On the other hand, an example of a procedural limitation would be to offer the DM a linear scale of 1 to 9 which does not cover all the possible values required to be cardinally consistent. Some of these limitations will be discussed in detail later (see also \citep{siraj2015contribution}). 

\begin{figure*}[tbh]
\begin{centering}
\includegraphics[width=0.85\textwidth]{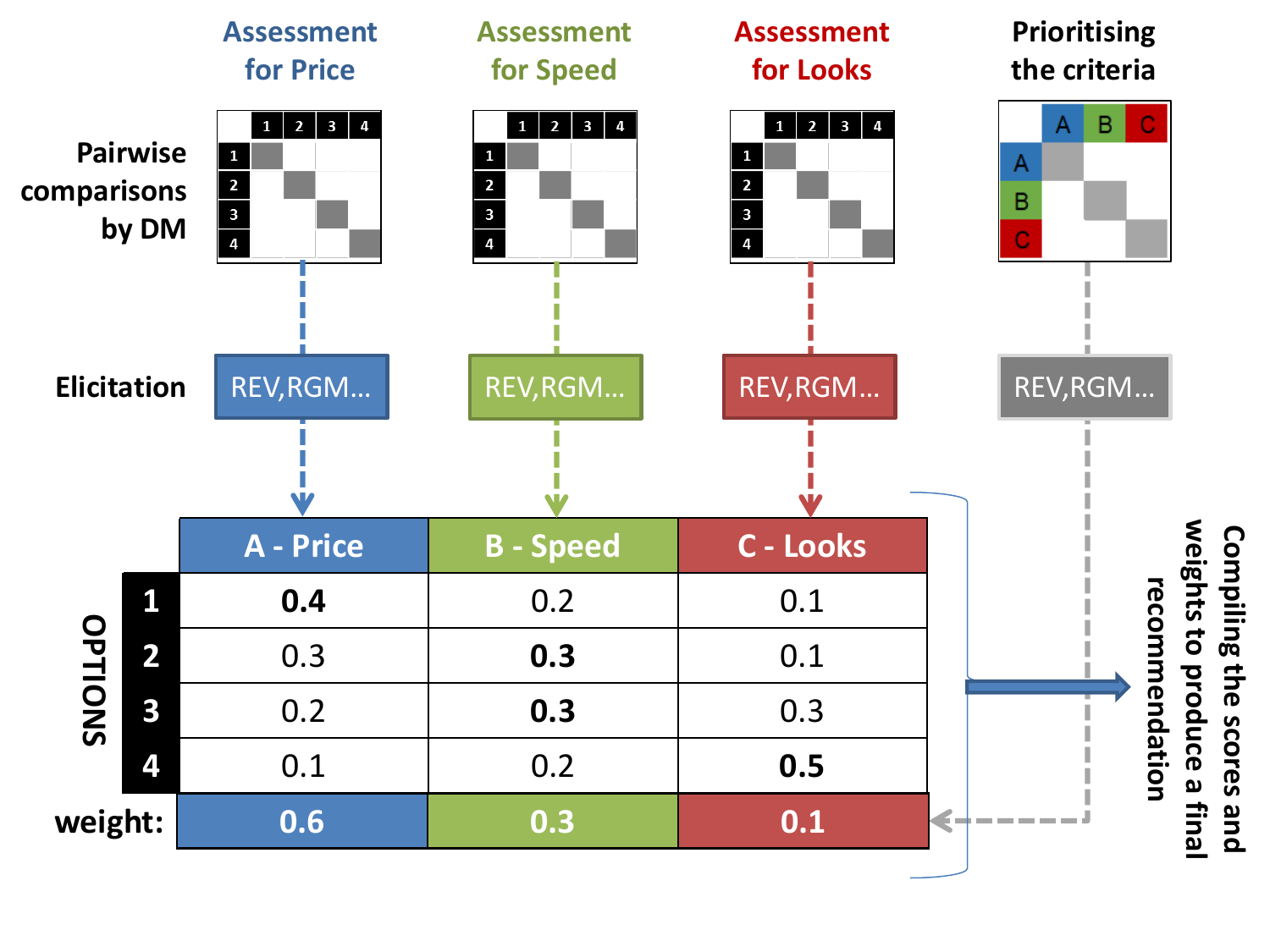} 
\par\end{centering}
\centering{}\caption{The PC matrix acquired for the top-level criteria\label{fig:typical-ahp-overview}}
\end{figure*}

In the case of inconsistency, eliciting the priority vector (i.e. weight vector or evaluation vector) is not a straight-forward task. There are numerous algorithms proposed to estimate the priority vector from an inconsistent set of judgements, however, it remains a widely debated issue as there is no single method that can be justified as the most appropriate method for eliciting preferences from inconsistent judgements. The most widely-used technique involving  pairwise comparison judgements is the Analytic Hierarchy Process (AHP) \citep{Saaty1980} where DMs structure their criteria into a hierarchy and then evaluate alternatives with respect to each of these criteria. Fig. \ref{fig:typical-ahp-overview} demonstrates this technique with the help of the car selection problem discussed earlier. In this figure, the DM has evaluated four alternatives with respect to each of Price, Speed and Looks (therefore, providing three sets of pairwise comparison judgements). Note that these judgements are shown in the form of a 4-by-4 matrix, which is a standard representation used to show pairwise comparisons. On the top-right of this figure, the DM has also provided the relative importance of the three criteria, again by using pairwise comparison judgements. 
These pairwise comparison judgements are then used to elicit priority vectors with the help of some elicitation technique. For example, the priority elicitation may be used to rank available options \citep{Sziklai2022,zhang2021personalized} or might be used in group decision making for eliciting weights of importance~\citep{fan2010approach,zhang2021personalized}. Although there are many techniques for priority elicitation, the most widely used techniques are Right Eigenvector (REV) \citep{Saaty1980} and Row Geometric Mean (RGM) \citep{Crawford1987}. These elicited evaluation priority vectors are then used to construct a decision table as shown at the bottom right of this figure (Fig. \ref{fig:typical-ahp-overview}). At the bottom of this table, note that the criteria weights (weight vector) can also be elicited using the same REV or RGM method. Finally, all these evaluation and weight vectors can be put together in a preference vector that can be used to generate some aggregated scores in order to produce some form of recommendations to the decision maker(s).

Historically, the REV method has been widely used for eliciting preference vectors for both consistent and inconsistent judgements. Although there exists a number of techniques to measure inconsistency \citep{jin2018novel}, it is most commonly measured in terms of the Consistency Ratio (CR) which is an Eigenvalue-based measure \citep{Saaty1980}. 

So far, we have assumed that the DMs provide a complete set of pairwise comparison judgements. However, in practice, it is quite common to confront situations where DMs provide an incomplete set of judgements, and where it is not always possible to ask DMs to provide the missing data \citep{li2022consensus,wan2022personalized}. This leaves us in situations where preferences should be elicited from inconsistent, as well as, incomplete sets of judgements. There exists a number of techniques to measure and analyse inconsistency in these incomplete sets of judgements as well \citep{Szybowski2020,Agoston2022}. 

The PC matrix is usually considered acceptable when the CR value remains below a value of 0.1. However, the REV method has been criticised due to its left-right eigenvector asymmetry, the use of arbitrary thresholds for inconsistency acceptability, as well as a few other further issues \citep{Brunelli2018}. Due to these shortcomings, several other methods have been proposed in the literature. A variety of these prioritisation methods are analysed and numerically compared in \citep{Choo2004} which concluded that there is no single best method that outperforms the others in every situation. Although REV is the most commonly used method, the RGM approach has gained popularity due to its mathematical properties, and due to its ease of implementation \citep{Kulakowski2020}. 

While focusing on this "single solution" aspect, it can be argued that the analysis of the inconsistency itself gets neglected. We contend that any prioritisation method must have the capabilities to focus on both aspects of the problem i.e. production of a "good quality" preference vector and also facilitation of an in-depth analysis. Whilst addressing and reducing the inconsistency is an important area, it cannot be completely removed in most cases. Therefore, we believe that it is also necessary to handle inconsistency with tools and techniques that can potentially unveil the sources of inconsistency in terms of plurality of mindsets in the same individual.

\subsection{Spanning tree approach}
\begin{figure*}[tbh]
\begin{centering}
\includegraphics[width=0.85\textwidth]{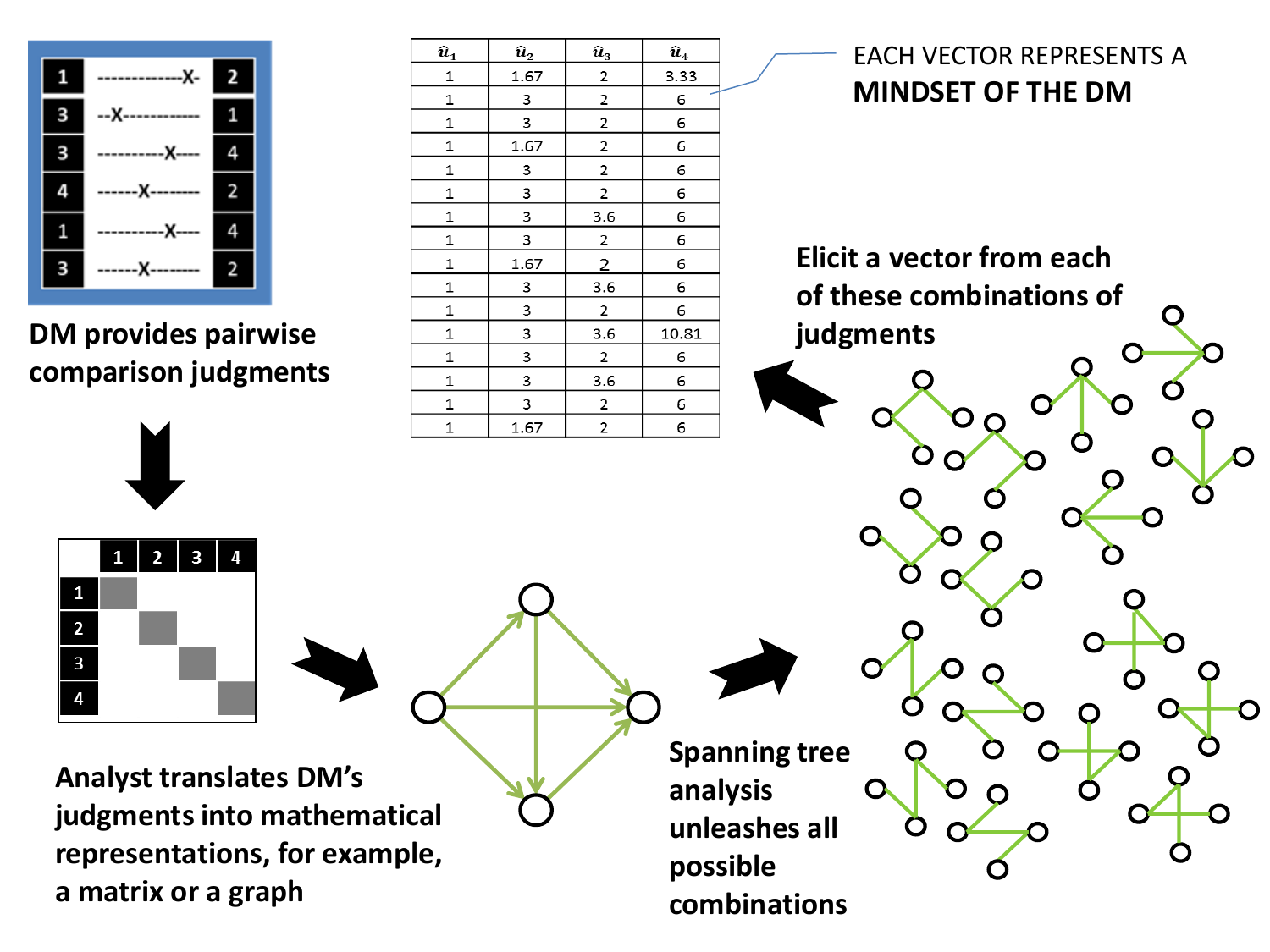} 
\par\end{centering}
\centering{}\caption{The PC matrix acquired for the top-level criteria\label{fig:spanning-tree-overview}}
\end{figure*}

In this context, a graph-theoretic approach was proposed to generate a set of all possible preference vectors through enumeration (see~\citep{Tsyganok2010}, and also \citep{Siraj2012}). This approach is briefly summarised in Fig. \ref{fig:spanning-tree-overview} where the DM provides a set of PC judgements (shown on the top-left) which is then translated into a fully connected graph (shown in the bottom-centre). This graph is then analysed using spanning tree analysis to identify all possible combinations of judgements (see the bottom right), and eventually, to generate all possible priority vectors (shown in the top-middle). Each of these alternative preference vectors essentially represents a mindset of the DM. 

The proposed method was shown to have a number of desirable properties, however, since the original method used the arithmetic mean to calculate the average of all these vectors, it was again focusing on the "single solution" aspect. 
Also, one may argue that this computationally expensive enumeration will be overkill when the judgements are (fully) consistent, as all vectors will have identical values. On the contrary, we argue that this rarely happens in real life; in the majority of cases, human judgements are found to be inconsistent, and therefore, unleashing these multiple mindsets can provide useful insights in practical decision making problems.

More recently, \citep{Lundy2017} proposed a geometric mean version of this spanning tree approach and established the mathematical equivalence of this approach with RGM. They also proposed the possibility of using statistical techniques to help the DM gain insights into inconsistency. In this paper, we develop this idea further and demonstrate its usefulness when we apply this technique at a meta-level, that is, applying it to the whole problem involving multiple criteria, instead of processing just one set of judgements.

\section{Stochastic approach to spanning trees\label{sec:spanning}}

We have seen that until now the multiplicity of preference vectors supplied by the spanning trees approach has been amalgamated into a single "representative" preference vector either using the arithmetic mean or geometric mean. However, with this approach  the information about the plurality of mindsets is lost. This is the price to pay to get a single overall ranking of alternatives. A different way of thinking could be to consider the plurality of ranking supplied by the plurality of preference vectors corresponding to the many combinations of (evaluation and weighting) spanning trees, that in turn correspond to the multiple mindsets.

\subsection{Problem formulation}
Consider a situation where the DM has $n$ alternative options to choose from, and $m$ criteria to consider whilst making this decision. These alternatives and criteria can be denoted as:
\begin{itemize}
\item Set of alternatives: $A=\{a_1,\ldots,a_i,\ldots,a_n\},$
\item Set of criteria: $G=\{g_1,\ldots,g_j,\ldots,g_m\}$
\end{itemize}

In order to assess these alternatives with respect to each criterion, $m$ sets of pairwise comparison judgements will be collected. Let us denote each of these pairwise comparison by
$$M^j=[c^j_{i_1,i_2}], g_j \in G$$
with $c^j_{i_1,i_2}$ being the pairwise comparison judgement of alternative $a_{i_1}$ with alternative  $a_{i_2}$ with respect to criterion $g_j$.

The DM also needs to assess the relative importance of the criteria, and therefore another pairwise comparison matrix will be required for elicitation, say:
$$M^G=[c^G_{j_1,j_2}]$$
with $c^G_{j_1,j_2}$ being the pairwise comparison judgement of criterion $g_{j_1}$ with criterion  $g_{j_2}$.

\subsubsection{Spanning trees}
Considering a generic pairwise comparison matrix $M=[c_{rs}]$, a spanning tree $\tau_{k}$ consists of $(n-1)$ mutually independent judgements out of the total judgements, that is   $$\tau_{k}=\{c^k_{r_1,s_1},\ldots,c^k_{r_{n-1},s_{n-1}}\}$$ with $c^k_{r_1,s_1},\ldots,c^k_{r_{n-1},s_{n-1}}$ independent judgements. Each of these spanning trees can be used to calculate a weight priority vector or an evaluation priority vector, and hence, we will interchangeably use the term "spanning tree" and "spanning tree vector".

For the matrices $M^j$ assessing alternatives with respect to the considered criteria $g_j, j=1,\ldots,m$, the set of spanning trees is denoted as:
$$\Tau^j=\{\tau^j_{k_r}\},$$

For the matrix $M^G$ assessing the importance of criteria, the set of spanning trees is denoted as:
$$\Tau^G=\{\tau_{k_r}^G\},$$

This gives us a total of $m+1$ sets of spanning trees (vectors), and therefore, $m+1$ sets of priority vectors. 

\begin{figure*}[tbh]
\begin{centering}
\includegraphics[width=0.75\textwidth]{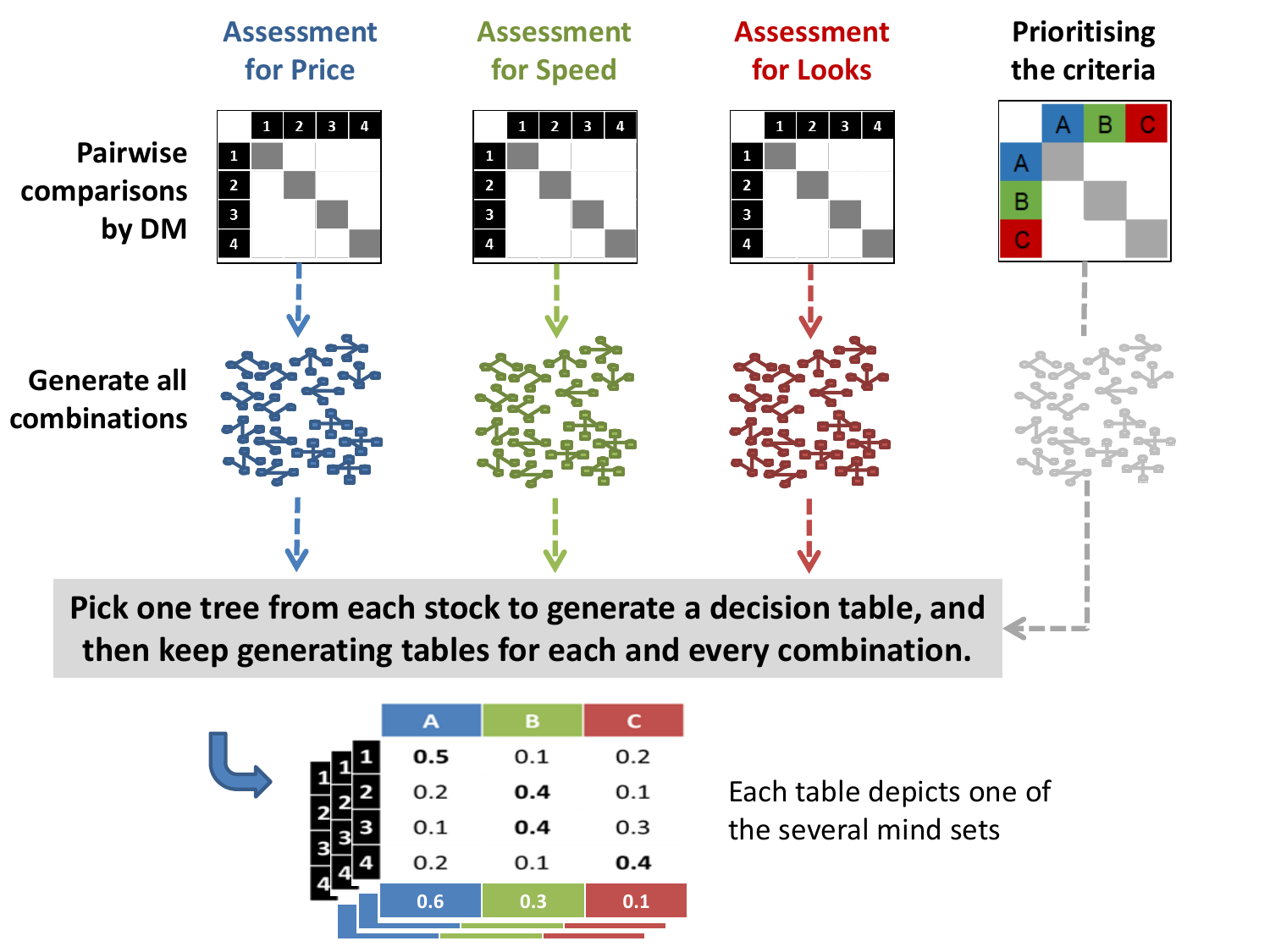} 
\par\end{centering}
\centering{}\caption{Creating a number of decision tables by using different combinations of spanning tree vectors \label{fig:revealing-all-preferences}}
\end{figure*}

As shown in Fig.  \ref{fig:revealing-all-preferences}, we can pick one tree from each of these sets and construct a decision table that is traditionally used to calculate the overall preferences (that is, the overall preference vector). A given combination of spanning trees can be denoted as
$$(\tau_{k_1}^1,\ldots,\tau_{k_m}^m,\tau_{k_G}^G) \in \Tau^1\times \ldots \Tau^m \times \Tau^G,$$

\subsubsection{Evaluating the combinations of spanning trees}
The evaluation vector of alternatives $a_i \in A$  corresponding to the  spanning tree $\tau^j_{k_j} \in \Tau^j$ can thus be represented as:
$$(u^{k_j}_j(a_1),\ldots,u^{k_j}_j(a_n))$$

while the weight vector corresponding to the  spanning tree $\tau^G_{k_G} \in \Tau^G$ can be represented as:
$$(w^{k_G}_1,\ldots,w^{k_G}_m)$$

Now the overall evaluation for alternative $a_i \in A$ given by the combination of spanning trees $\bm{\tau}=$\\
${\scriptstyle (\tau_{k_1}^1,\ldots,\tau_{k_m}^m,\tau_{k_G}^G) \in \Tau^1\times \ldots \Tau^m \times \Tau^G}$ will be:

$$u_{\bm{\tau}}(a_i)=w^{{k_G}}_1 u^{k_1}_1(a_i)+ \ldots+w^{k_G}_m u^{k_m}_m(a_i)$$

Using this overall evaluation, we can define the set of combinations of spanning trees 
for which, for any $a_{i_1},a_{i_2} \in A$
\begin{itemize}
    \item $a_{i1}$ is preferred to $a_{i2}$: \\
$${ B(a_{i_1} \succ a_{i_2})=\{\bm{\tau} \in \scriptstyle \Tau^1\times \ldots \Tau^m \times \Tau^G :u_{\bm{\tau}}(a_{i_1})>u_{\bm{\tau}}(a_{i_2}) \textstyle \} }$$

    \item $a_{i1}$ is indifferent to $a_{i2}$: \\
$${ B(a_{i_1} \sim a_{i_2})=\{\bm{\tau}\in \scriptstyle \Tau^1\times \ldots \Tau^m \times \Tau^G :u_{\bm{\tau}}(a_{i_1})=u_{\bm{\tau}}(a_{i_2}) \textstyle \} }$$

\end{itemize}

Furthermore, the set of combinations of spanning trees $\bm{\tau} \in \scriptstyle \Tau^1\times \ldots \Tau^m \times \Tau^G$ for which alternative $a_i \in A$ attains the $p-th$ position  with respect to the overall evaluation $u_{\bm{\tau}}(a_{i})$ can be formulated as:
$$R(a_i,p)=\{\bm{\tau} \in \scriptstyle \Tau^1\times \ldots \Tau^m \times \Tau^G:rank(a_i,\bm{\tau})=p\textstyle \}$$
with 
$$rank(a_i,\bm{\tau})=1+\sum_{i^\prime \neq i}\rho(u_{\bm{\tau}}(a_{i^\prime})>u_{\bm{\tau}}(a_{i}))$$
where $\rho(false)=0$ and $\rho(true)=1$. 

Inspired by SMAA \citep{lahdelma1998smaa,lahdelma2001smaa}, the probability that alternative $a_{i_1}$ is preferred to alternative $a_{i_2}$, with $a_{i1},a_{i2}\in A$ can therefore be represented as:
$$P(a_{i1} \succ a_{i2})=\frac{card(B(a_{i1} \succ a_{i2}))}{card(\scriptstyle \Tau^1\times \ldots \Tau^m \times \Tau^G \textstyle)}$$ 

Remembering that the total number of spanning trees in a graph of $k$ nodes is $k^{k-2}$ \citep{cayley1889theorem}, and, consequently $card(\scriptstyle \Tau^1\times \ldots \Tau^m \times \Tau^G \textstyle)=m^{m-2}n^{m(n-2)}$, $P(a_{i1} \succ a_{i2})$ becomes
$$P(a_{i1} \succ a_{i2})=\frac{card(B(a_{i1} \succ a_{i2}))}{m^{m-2}n^{m(n-2)}},$$

Analogously, the probability that alternative $a_{i} \in A$ attains the $p-th$ ranking position is given by:
$$P(rank(a_i,\bm{\tau})=p)=\frac{R(a_i,p)}{card(\scriptstyle \Tau^1\times \ldots \Tau^m \times \Tau^G \textstyle)}$$
$$=\frac{R(a_i,p)}{m^{m-2}n^{m(n-2)}},$$ 

The probability that an alternative $a$ is preferred to another alternative $b$ can be termed the Pairwise Winning Index (PWI), and the probability that an alternative attains a given ranking position can be termed the Rank Acceptability Index (RAI).

\subsection{Spanning trees for incomplete pairwise comparison judgements}

Harker \citep{Harker1987a} investigated incomplete sets of judgements where the DMs are allowed to respond with "do not know" or "not sure" to some judgements. This is an important issue to investigate as the probability of acquiring an incomplete set of PC judgements increases with an increase in the total number of items for comparison \citep{Fedrizzi2007}. Both the REV and the RGM methods are inappropriate in such cases due to the fact that the PC matrix cannot be constructed without estimating/imputing the missing judgements. 

The spanning trees approach can be applied also to partial comparison matrices with the help of the Kirchoff formula (See details in \citep{Siraj2012}). This implies that generating combinations of spanning trees is not restricted to only complete sets of judgements, and therefore, generating preference frequencies and rank-order frequencies are possible from incomplete pairwise comparisons, without any modification. 

The implication of having an incomplete set of judgements is that the total number of combinations will be smaller than the number possible from a complete sets of judgements. This permits the DM to only express values for those PC judgements for which he/she is willing to provide input, avoiding forcing the DM to provide complete set of judgements. This ensures maintaining the integrity and reliability of the information considered in the decision support process \citep{li2022consensus,zhang2021personalized}. 

\subsection{Didactic example}

Let us show how our approach works in practice by using the classic example of school selection proposed in \citep{Saaty1980}. The parents have to decide the high school for their son. They consider six criteria being the following
\begin{enumerate}
\item $g_1$: Learning,
\item $g_2$: Friends,
\item $g_3$: School life,
\item $g_4$: Vocational training,
\item $g_5$: College preparation,
\item $g_6$: Music classes.
\end{enumerate}
There are three alternatives corresponding to three schools denoted $A$, $B$ and $C$. 
For each criterion, the parents compared in pairs the schools as shown in the following pairwise comparison matrices.

\begin{equation}
M_{g_1}=
\begin{pmatrix}
    3 & \frac{1}{3} & \frac{1}{2} \\
    3 & 1 & 3  \\
    2 & \frac{1}{3} & 1 \\
  \end{pmatrix}
  \label{M_g1}
\end{equation}

\begin{equation}
M_{g_2}=
\begin{pmatrix}
    1 & 1 & 1 \\
    1 & 1 & 1  \\
    1 & 1 & 1 \\
  \end{pmatrix}
  \label{M_g2}
\end{equation}

\begin{equation}
M_{g_3}=
\begin{pmatrix}
    1 & 5 & 1 \\
    \frac{1}{5} & 1 &  \frac{1}{5}  \\
    1 & 5 & 1 \\
  \end{pmatrix}
  \label{M_g3}
\end{equation}

\begin{equation}
M_{g_4}=
\begin{pmatrix}
    1 & 9 & 7 \\
    \frac{1}{9} & 1 & \frac{1}{5}  \\
    \frac{1}{7} & 5 & 1 \\
  \end{pmatrix}
  \label{M_g4}
\end{equation}

\begin{equation}
M_{g_5}=
\begin{pmatrix}
    1 & \frac{1}{2} & 1 \\
    2 & 1 & 2  \\
    1 & \frac{1}{2} & 1 \\
  \end{pmatrix}
  \label{M_g5}
\end{equation}

\begin{equation}
M_{g_6}=
\begin{pmatrix}
    1 & 6 & 4 \\
    \frac{1}{6} & 1 & \frac{1}{3}  \\
    \frac{1}{4} & 3 & 1 \\
  \end{pmatrix}
  \label{M_g6}
\end{equation}

After comparing the alternatives with respect to these criteria, the parents also compared the criteria in terms of their importance as shown below.

\begin{equation}
M_{criteria}=
\begin{pmatrix}
    1 & 4 & 3 & 1 & 3 & 4 \\
    \frac{1}{4} & 1 & 7 & 3 & \frac{1}{5} & 1 \\
    \frac{1}{3} & \frac{1}{7} & 1 & \frac{1}{5} & \frac{1}{5} & \frac{1}{6} \\
    1 & \frac{1}{3} & 5 & 1 & 1 & \frac{1}{3} \\
    \frac{1}{3} & 5 & 5 & 1 & 1 & 3 \\
    \frac{1}{4} & 1 & 6 & 3 & \frac{1}{3} & 1 \\
  \end{pmatrix}
  \label{M_criteria}
\end{equation}

The consistency ratios for the above pairwise comparison matrices are as follows:
\begin{itemize}
\item CR($M_{g_1}$) = 0.04,
\item CR($M_{g_2}$) = 0,
\item CR($M_{g_3}$) = 0,
\item CR($M_{g_4}$) = 0.18,
\item CR($M_{g_5}$) = 0,
\item CR($M_{g_6}$) = 0.04,
\item CR($M_{criteria}$) = 0.24
\end{itemize}

As we can see, there are four matrices which are not completely consistent and two of these, namely $M_{g_4}$ and $M_{criteria}$ have unacceptable levels of inconsistency as measured by the usual threshold of 0.1 for the CR value \citep{Saaty1980}.

The priorities $u_j(X), X=A,B,C, j=1,\ldots,6$, can be calculated (using the RGM method) from the pairwise comparisons matrices $M_{g_1}-M_{g_6}$ which represent the evaluations of schools with respect to the considered criteria. These priorities are shown in Table \ref{table:tbl-Scores} for all three schools. The weights of the considered criteria can also be obtained (using the RGM method) from pairwise comparison's matrix $M_{criteria}$, which are the following:
$w_1=0.32$, $w_2=0.14$, $w_3=0.03$, $w_4=0.13$, $w_5=0.24$, $w_6=0.14$.

\begin{table}[htb]
    \centering 
    \caption{Scores for alternatives with respect to each criterion}
    \begin{tabular}{ lllllllll}
      \hline
        Alternatives & $g_1$ & $g_2$ & $g_3$ & $g_4$ & $g_5$ & $g_6$  \\ \hline
        School A & 0.16 & 0.33 & 0.45 & 0.77 & 0.25 & 0.69  \\ \hline
        School B & 0.59 & 0.33 & 0.09 & 0.05 & 0.50 & 0.09  \\ \hline
        School C & 0.25 & 0.33 & 0.46 & 0.17 & 0.25 & 0.22  \\ 
      \hline
    \end{tabular}
	\label{table:tbl-Scores}
\end{table}

Using the priorities $w_j$ and $u_j(X), X=A,B,C, j=1,\ldots,6$, we can compute the overall evaluation $u(X)$ of each school, with
$$u(X)=\sum_{j=1}^6 w_j u_j(X)$$
obtaining the following results:
$$u(A)=0.37, u(B)=0.38, u(C)=0.25.$$ 

Let us now handle the same problem with the spanning tree approach. With this aim, we have to consider all the combinations of spanning trees from the pairwise comparison matrix of criteria $M_{criteria}$ and from the pairwise comparison matrices of alternatives with respect to the considered criteria $M_{g_j}, j=1,\ldots,m$. Remembering that with $m$ criteria and $n$ alternatives, we have ${m^{(m-2)}}{n^{(n-2)m}}$ combinations of spanning trees, for the decision problem at hand, we have  $6^4\cdot 3^6 = 944784$ combinations of spanning trees. We compute weights of criteria 
$$w^k_1,\ldots,w^k_6$$
and evaluations of alternatives with respect to the considered criteria 
$$u^k_j(A),u^k_j(B),u^k_j(C)), j=1,\ldots,6,$$
for each combination of spanning trees 
$\bm{\tau}=(\tau_{k_1}^1,\ldots,\tau_{k_6}^m,\tau_{k_G}^G) \in \Tau^1\times \ldots \Tau^6 \times \Tau^G$.

\begin{figure*}[tbh]
\begin{centering}
\includegraphics[width=0.75\textwidth]{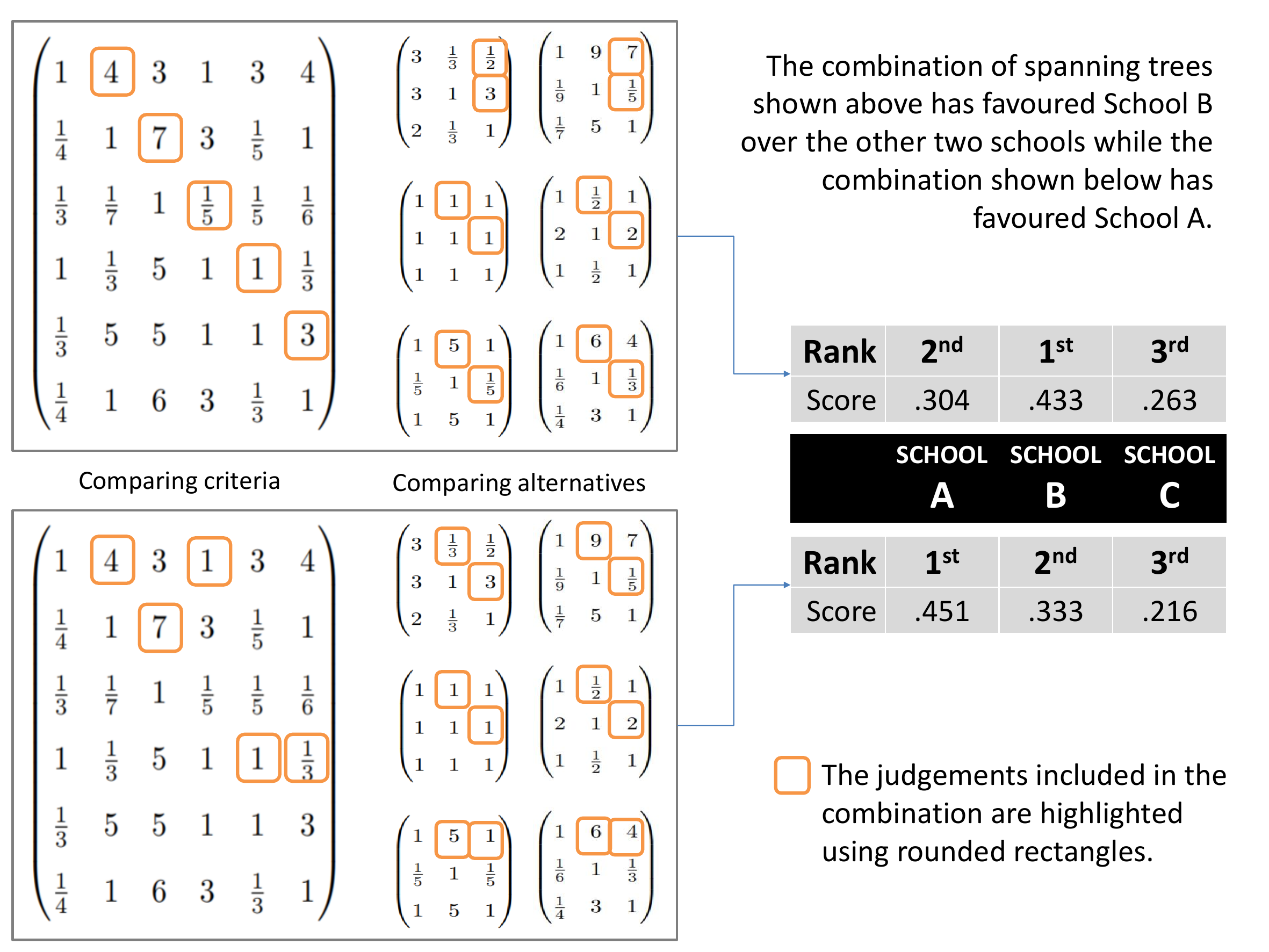} 
\par\end{centering}
\centering{}\caption{Demonstrating the two different combinations of spanning trees, and the difference in their rankings and scores. \label{fig:didactic-example-school}}
\end{figure*}

Considering all the combinations of spanning trees from $\Tau^1\times \ldots \Tau^6 \times \Tau^G$ we can compute 
\begin{itemize}
\item Table \ref{tab:PWI} which for each pair of alternatives shows the PWI, that is the probability that the alternative in the row is preferred to the alternative in the column (the figure in parenthesis shows the total number of combinations of spanning tress for which the alternatives attain the ranking position, indicated by the row label).
\item Table \ref{tab:RAI} which shows the RAIs which are the probabilities that each alternative attains a given rank (the figure in parenthesis shows the total number of combinations of spanning trees for which the alternative on the row is preferred to the alternative on column),
\end{itemize}

The results of Table \ref{tab:PWI} and Table \ref{tab:RAI} can be interpreted in terms of plurality of mindsets. For example, considering School B and School C in Table \ref{tab:PWI}, the plurality of mindsets induce the probability of 91\% that School A is preferred to School C, and a probability of 89\% that School B is preferred over School C. Also, the first row of Table \ref{tab:RAI} suggests that the DM is in two mind sets for which School A and School B are considered best, and the relevance of the two mindsets can be measured by their respective RAIs of 51\% and 49\%. On the other hand, School C has an RAI of 80\% to be at the least favourable position in ranking. In other words, there is a wealth of information and interpretation that can be deduced from the PWI and RAI that would otherwise be missing from an approach which focuses solely on the production of the final preference vector alone.

\begin{table}[htb]
    \centering 
    \caption{PWIs in terms of preference frequencies over 944784 combinations of spanning trees}
    \label{tab:PWI}
    \begin{tabular}{llll}
      \hline
         		& School A & School B & School C   \\ \hline
        School A & X &  0.51 (483246) & 0.91 (855063)  \\ \hline
        School B & 0.49 (461538) & X & 0.89 (842130)   \\ \hline
        School C &  0.09 (89721) & 0.11 (102654) & X   \\ \hline
      \hline
    \end{tabular}

\end{table}

\begin{table}[htb]
    \centering 
    \caption{RAIs in terms of rank order frequencies over 944784 combinations of spanning trees}
    \label{tab:RAI}
    \begin{tabular}{ llll}
    \hline
       		& School A & School B & School C   \\ \hline
        $1^{st}$ & 0.51 (483084) & 0.49 (461268) & 0.00 (432)  \\ \hline
        $2^{nd}$ & 0.49 (372141) & 0.40 (381132) & 0.11 (191511)   \\ \hline
        $3^{rd}$ & 0.09 (89559) & 0.11 (102384) & 0.80 (752841)    \\ \hline
    \end{tabular}
\end{table}

\section{Random spanning trees} \label{sec:randomspanning}
We have seen that, for a problem with $n$ alternatives evaluated across $m$ criteria, the total number of combinations of spanning trees increases exponentially with the problem size - making it impractical to calculate the exact probabilities $P(rank(a_i,\bm{\tau})=p)$ and $P(a_{i1} \succ a_{i2})$ for large problems.

We therefore propose using an approach which, combined with statistical sampling theory, allows us to provide estimates of the required probabilities  $P(rank(a_i,\bm{\tau})=p)$ and $P(a_{i1} \succ a_{i2})$ to within any user defined degree of accuracy according to any user defined level of confidence.

Of course we cannot physically select a statistical random sample from the population of spanning tree combinations since this would require the generation of the population itself and this is exactly what we aim to avoid. We can however use the 'random walk' procedure to generate a tree and hence a sample of trees (see  \citep{Aldous1990,Broder1989}. Indeed, the 'random walk' procedure has been proven to generate a true statistical random sample in the sense that the generated sample is equivalent to selection of a statistical random sample from the population, that is, where each tree has the same uniform probability of being selected. See  \citep{Broder1989,Aldous1990} for more details.

As we discussed previously, the possibility of having incomplete sets of pairwise comparison judgements will arise in practice and so it is important to note that the concept of random walks is equally applicable to these incomplete sets without modification. 

\begin{figure*}[tbh]
\begin{centering}
\includegraphics[width=0.75\textwidth]{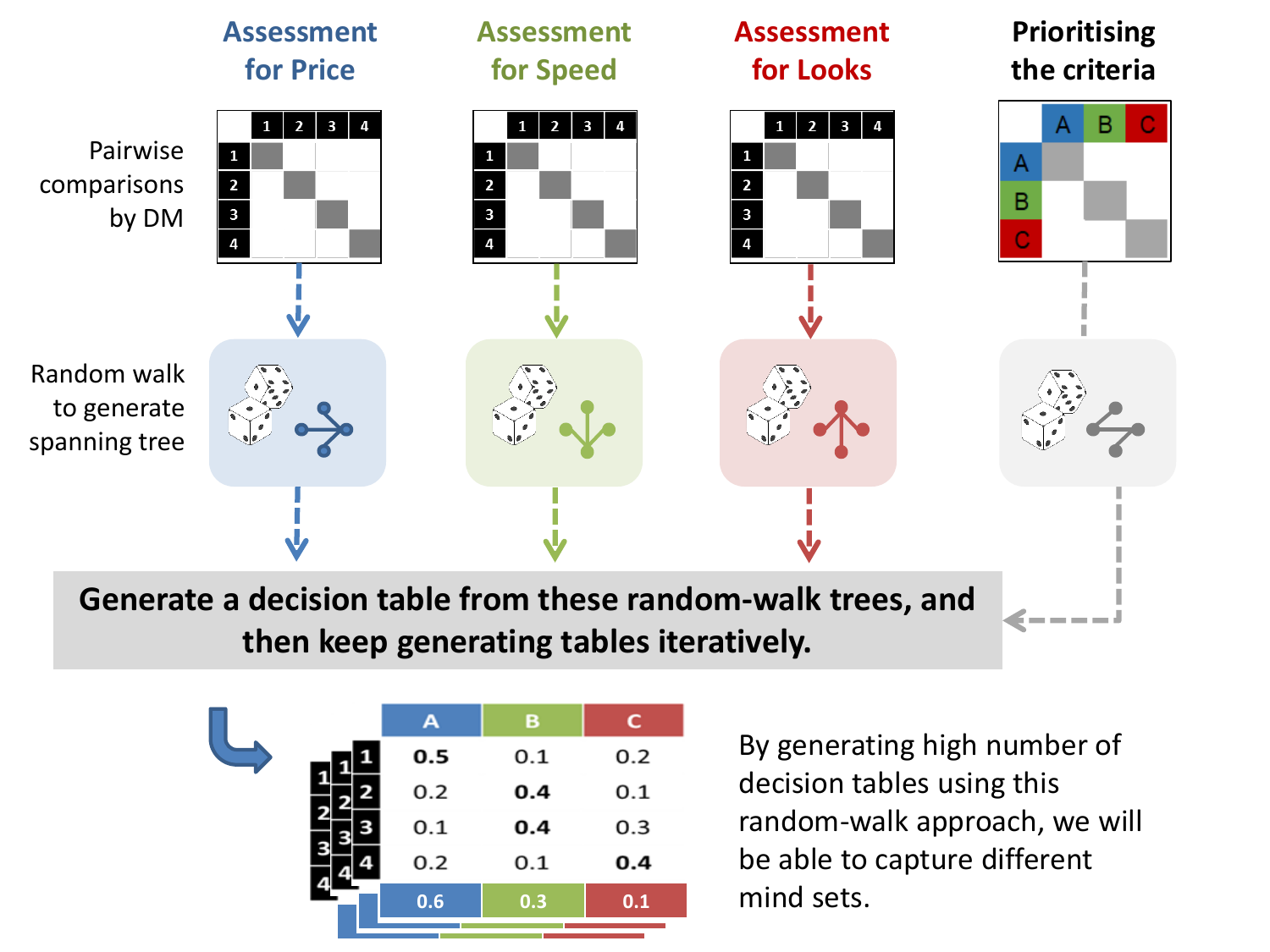} 
\par\end{centering}
\centering{}\caption{The PC matrix acquired for the top-level criteria\label{fig:forest-with-random-walk}}
\end{figure*}

Since each iteration of the procedure generates a new member of the random sample, the total number of iterations used in the procedure is equivalent to sample size and we can use statistical large sample theory to determine the number of iterations required to generate sample parameters (including the probabilities $P(rank(a_i,\bm{\tau})=p)$ and  $P(a_{i1} \succ a_{i2})$) to any specified degree of accuracy and with any specified level of confidence.

That is, if we want the estimated probability to have an accuracy within $\lambda$ and with a $C$ percent level of confidence, then the required number of iterations would be \citep{Tervonen2007}:

$$It(\lambda,C) = \frac{ {Z_C}^2 }{4\lambda^2}$$ where $Z_C$ is the z-score calculated from the standardised normal distribution curve.

For example, if we want to achieve an accuracy within 0.01 and with a $99\%$ level of confidence, then the required number of iterations would be
$$It(0.01,99\%) = \frac{ Z_{99}^2 }{4 \times 0.01^2}$$

As the Z value for a 99\% confidence interval is equal to 2.58, 
$$It(0.01,99\%) = \frac{ {(2.58)}^2 }{4 \times 0.01^2} = 16,641$$

This suggests that we need to perform more than 16,641 iterations if we want to achieve an accuracy within $\pm0.01$ with $99\%$ confidence.

\subsection{Applying random walk procedure to the school example}

In the example below, we illustrate the use of random walks by repeating an experiment multiple times where each experiment itself uses the required number of iterations $It$ determined by the formula above. We do this on the previously discussed School example as this is a tractable problem where the whole population of solutions is available. Later, we will also extend this idea to a bigger problem where generating the whole population is impractical (see Section \ref{sec:telecomstudy}).

Table \ref{tbl:school_dominance_randomwalk} shows the PWIs obtained through random walks (see the table on left), as well as the PWIs obtained by selecting a sample from the whole population of solutions (see the table on right). The values shown in these tables are the average PWIs obtained from 20 experiments. Recall that the whole population was generated previously by enumerating all possible combinations of spanning trees as was shown in Table \ref{tab:PWI} earlier. Examining the figures in Tables \ref{tbl:school_dominance_randomwalk}, we can see that the estimated PWI values from random walks and from the selected sample both lie within 1\% of the population values that were previously shown in Table \ref{tab:PWI}. 

\begin{table*}[!htb]
    \caption{PWIs based on preference frequencies of the school example}
	\begin{tabular}[width=0.9\pagewidth]{cc}
	    \begin{minipage}{.5\linewidth}
	        \begin{tabular}{llll}
				\multicolumn{4}{c}{From random walks}\\ \hline
		         		& School A & School B & School C   \\ \hline
		        School A & - & 51.3\% & 90.5\%  \\ \hline
		        School B & 48.7\% & - & 89.2\%   \\ \hline
		        School C & 9.5\% & 10.8\% & -   \\ \hline
	        \end{tabular}
	    \end{minipage} &
	
	    \begin{minipage}{.5\linewidth}
	        \begin{tabular}{llll}
				\multicolumn{4}{c}{Sampling from population}\\ \hline
		         		& School A & School B & School C   \\ \hline
		        School A & - & 50.2\% & 90.1\%  \\ \hline
		        School B & 49.8\% & - & 88.6\%   \\ \hline
		        School C & 9.9\% & 11.4\% & -   \\ \hline
	        \end{tabular}
	    \end{minipage} 
		\label{tbl:school_dominance_randomwalk}
	\end{tabular}

\end{table*}

\begin{table*}[]
   \caption{RAIs based on rank order profiles generated for the school example}
	\begin{tabular}{cccc}
	    \begin{minipage}{.5\linewidth}
	        \begin{tabular}{llll}
				\multicolumn{4}{c}{From random walks}\\ \hline
		         		& School A & School B & School C   \\ \hline
		        $1^{st}$ & 51.2\% & 48.7\% & 0.0\%  \\ \hline
		        $2^{nd}$ & 39.3\% & 40.5\% & 20.2\%   \\ \hline
		        $3^{rd}$ & 9.5\% & 10.8\% & 79.7\%    \\ \hline
	        \end{tabular}
	    \end{minipage} &
	
	    \begin{minipage}{.5\linewidth}
	        \begin{tabular}{llll}
				\multicolumn{4}{c}{Sampling from population}\\ \hline
		         		& School A & School B & School C   \\ \hline
	        $1^{st}$ & 50.2\% & 49.8\% & 0.0\%  \\ \hline
	        $2^{nd}$ & 39.9\% & 38.8\% & 21.3\%   \\ \hline
	        $3^{rd}$ & 9.9\% & 11.4\% & 78.7\%    \\ \hline
	        \end{tabular}
	    \end{minipage} 
	\label{tbl:school_ranking_randomwalk}
	\end{tabular}
\end{table*}

We also performed the same analysis for the RAIs, as summarised in Table \ref{tbl:school_ranking_randomwalk}. The table on the left side shows that the rank-order frequencies obtained through random walks also lie within 1\% of the population values shown previously in Table \ref{tab:RAI}. The rank-order frequencies obtained by selecting a sample are shown on the right side of this table. As with the PWI, this also illustrates that the accuracy and confidence level results for the random walk sample echo the same results had we been able to physically select a statistical random sample from the population of spanning tree combinations.

\section{Telecom backbone example}\label{sec:telecomstudy}
In order to demonstrate the use of spanning trees to explore multiple mindsets, we consider the practical data acquired in a recent study: the selection
of a backbone infrastructure for telecommunication in rural areas \citep{Gasiea2010}. This study focused on the use of a structured decision making approach towards selecting the telecommunication infrastructure for the rural areas of developing countries. The lack of adequate telecommunications infrastructure in these parts of the world remains a major obstacle for providing affordable services.

The study considered four options of Fiber-optic cable (G1), Power-line communication (G2), Microwave link (G3) and Satellite communication (G4). The authors of this study used AHP and Analytic Network Process (ANP) to structure the problem and acquired the necessary data on preferences and assessments from key stakeholders. The criteria used to compare these alternatives were grouped into six major categories including (A) technical, (B) infrastructural, (C)
economic, (D) social, (E) regulatory and (F) environmental factors. 

\begin{figure}[tbh]
\begin{centering}
\includegraphics[width=0.48\textwidth]{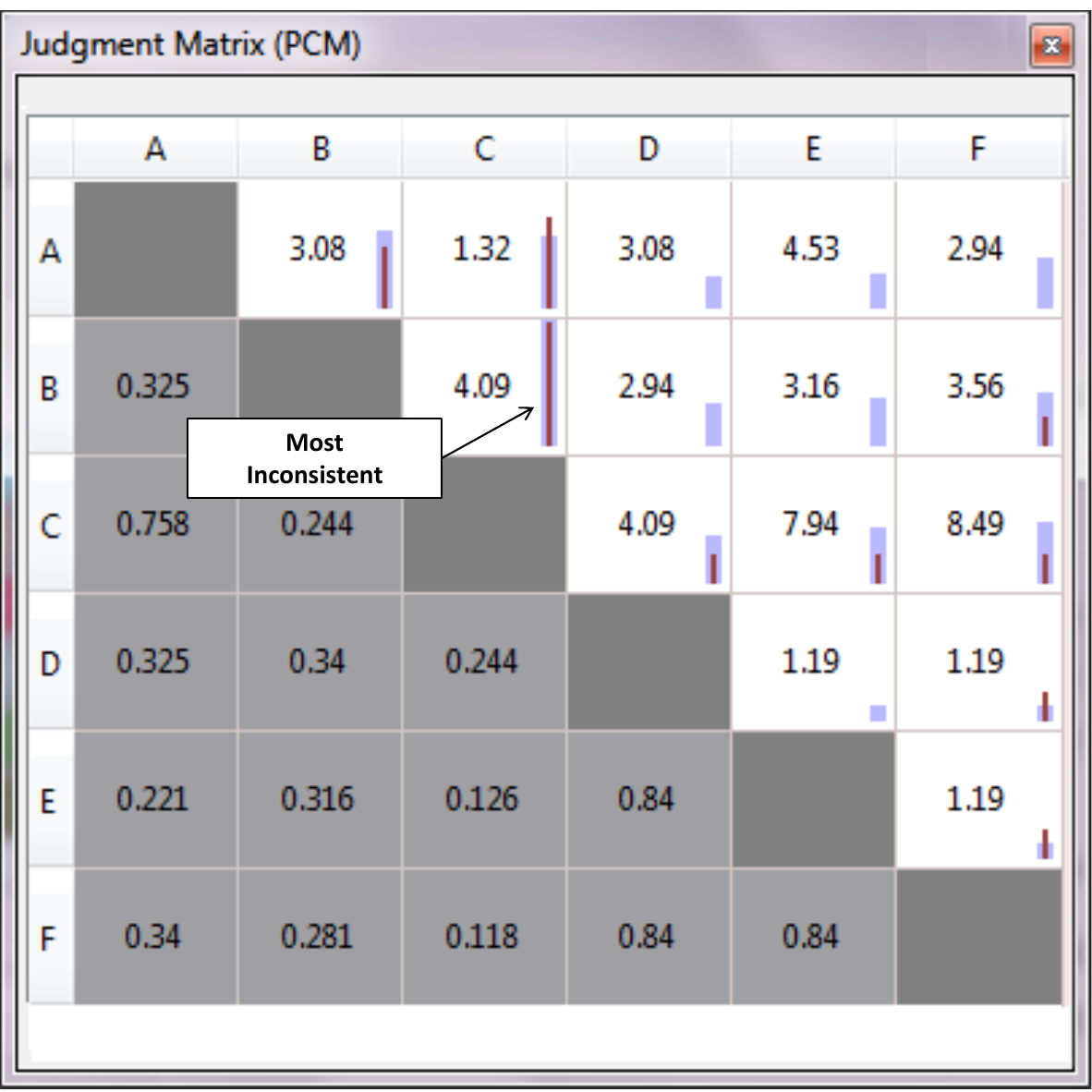} 
\par\end{centering}
\centering{}\caption{The PC matrix acquired for the top-level criteria\label{fig:criteria-top}}
\end{figure}

The PC matrix, $A_{top}$, acquired for prioritising these six categories (top-level criteria) is shown in Fig. \ref{fig:criteria-top}. Although $A_{top}$ is a transitive PC matrix, the estimated vector produced by the widely used REV method does not preserve the original order of preferences in these judgements.

The final weights calculated using the REV and RGM methods are found to be almost identical, as given in Table \ref{tab:Estimated-weights-final} in normalised form. Satellite communication (G4) is considered the
most preferred alternative with a weight of 29.95\% (using REV), followed by Microwave (G3) with a weight around 28.34\% (using REV). 

\begin{table}[tbh]
\centering{}%
\begin{tabular}{rcccc}
\hline 
 & Fiber  & Powerline  & Microwave  & Satellite\tabularnewline
 & $w_{G1}$  & $w_{G2}$  & $w_{G3}$  & $w_{G4}$\tabularnewline
\hline 
REV:  & $21.7\%$  & $20.1\%$  & $28.3\%$  & $29.9\%$\tabularnewline
RGM:  & $21.7\%$  & $20.1\%$  & $28.4\%$  & $29.8\%$\tabularnewline
\hline 
\end{tabular}\caption{Estimated weights for the available backbone infrastructure options\label{tab:Estimated-weights-final}}
\end{table}

Most criteria lie under the \emph{Technical} and \emph{Infrastructure} categories. The \emph{Technical} category includes nine criteria whilst the \emph{Infrastructure} category has eight criteria used to compare the alternatives. The two PC matrices for the \emph{Technical} and \emph{Infrastructure}
categories, $A_{tech}$ and $A_{infra}$, have been found to be intransitive and should be investigated along with $A_{top}$ for their impact on the final result.
\newcommand{\q}[1]{``#1''}
\subsection{Spanning trees analysis using Random walks}
In this case study, there are four alternatives and six top-level criteria, therefore, the number of \q{spanning-trees} solutions for the criteria matrix is $m^{(m-2)}$ = $6^4$ = 1296, and the number of  \q{spanning-trees} solutions for each alternative matrix is $n^{(n-2)}$ = $4^2$ = 16. As we pick one  \q{spanning-trees} solution from each matrix, there are $m^{(m-2)}n^{m(n-2)}$ = $1296\times16^6$ = $21,743,271,936$ combinations possible. Generating these billions of combinations is computationally expensive, so sampling is the logical way to estimate probabilities. Therefore we generated a sample of 20,000 solutions with the use of random walks. We generated 20 such samples in order to check whether all these samples produce similar results. In practice, only one sample should suffice but we took this approach for the purpose of investigation. 

The average scores for the four alternatives are given in Table \ref{tab:randomwalks-iterations}. Each row in this table represents one experiment of generating a sample of 20,000 solutions. 

\begin{table*}[tbh]
\centering{}%
\caption{Scores calculated using random walks; each experiment shows an average of 20k solutions\label{tab:randomwalks-iterations}}

    \begin{tabular}{lllll}
    \hline 
    Experiment \# & Fiber  & Powerline  & Microwave  & Satellite 
    \tabularnewline
    \hline 
    1         & 0.218295 & 0.193467 & 0.292819 & 0.29542  \\
    2         & 0.218428 & 0.193078 & 0.292313 & 0.29618  \\
    3         & 0.218372 & 0.192959 & 0.293072 & 0.295597 \\
    4         & 0.218190 & 0.193914 & 0.292752 & 0.295143 \\
    5         & 0.218980 & 0.19307  & 0.291989 & 0.295961 \\
    6         & 0.219305 & 0.193273 & 0.292496 & 0.294927 \\
    7         & 0.218776 & 0.192261 & 0.292680 & 0.296283 \\
    8         & 0.218472 & 0.194001 & 0.292665 & 0.294861 \\
    9         & 0.218751 & 0.193282 & 0.293100 & 0.294867 \\
    10        & 0.218287 & 0.193674 & 0.292790 & 0.295249 \\
    11        & 0.218830 & 0.193428 & 0.293161 & 0.294581 \\
    12        & 0.218442 & 0.193236 & 0.292886 & 0.295436 \\
    13        & 0.218468 & 0.193785 & 0.292581 & 0.295165 \\
    14        & 0.219373 & 0.192499 & 0.292469 & 0.295659 \\
    15        & 0.219038 & 0.193186 & 0.292143 & 0.295633 \\
    16        & 0.218294 & 0.194367 & 0.292860 & 0.29448  \\
    17        & 0.218691 & 0.193434 & 0.292697 & 0.295179 \\
    18        & 0.218749 & 0.193219 & 0.292860 & 0.295171 \\
    19        & 0.218288 & 0.194101 & 0.292734 & 0.294876 \\
    20        & 0.218517 & 0.192994 & 0.292892 & 0.295597
    \tabularnewline
    \hline 
    \end{tabular}

\end{table*}

Unlike the School example where we calculated frequencies by generating all the combinations of spanning trees, here we will calculate the frequencies from the samples generated by random walks. It can be argued that these frequencies may vary as they are based on randomly generated set of solutions. However, the aim here is to gain statistical insights, and therefore, these findings should remain useful as long as they produce reliable statistical findings. We investigated this reliability by repeating the same experiment twenty times and comparing the preference frequencies and rank-order frequencies. The standard deviations of these values do not exceed 0.005, which supports the argument that these stochastic results are quite reliable.

Table \ref{tab:preference-freq} shows the PWIs for all possible pairs. Looking at the first row, we can say that Fiber is preferred over Powerline in 63\% (see $0.63\pm0.004$ under the Powerline column) of the random walk solutions, however, it is seldom preferred over Microwave and Satellite. Looking at the second row, Powerline is dominated by all other choices. On the contrary, Microwave is clearly preferred over Fiber (with a probability of 87\%) and Powerline (with a probability of 96\%), however, it has been marginally preferred over Satellite (with a probability of 51\%). Looking at the last row, Satellite is also clearly preferred over Fiber (91\%) and Powerline (91\%) but not against Microwave. Therefore, we can argue that the Fiber and Powerline are the two clearly dominated alternatives, while the Microwave and Satellite are the two dominating alternatives. However, when comparing Microwave and Satellite with each other, Microwave is slightly more preferred but there is no clear winner. We have provided the standard deviations for each of these values to show that these scores didn't deviate much, as discussed earlier.

\begin{table*}[tbh]
\centering{}%
\caption{PWIs for Telecom backbone selection\label{tab:preference-freq}}
\begin{tabular}{rcccc}
\hline 
 & Fiber  & Powerline  & Microwave  & Satellite\tabularnewline
\hline 
Fiber:        & $-$  & $0.63\pm0.004$  & $0.13\pm0.002$  & $0.09\pm0.003$\tabularnewline
Powerline:  & $0.37\pm0.004$  & $-$ & $0.04\pm0.002$  & $0.09\pm0.002$\tabularnewline
Microwave: & \boldmath{$0.87\pm0.002$}  & \boldmath{$0.96\pm0.002$}  & $-$  & $0.51\pm0.004$\tabularnewline
Satellite:    & \boldmath{$0.91\pm0.003$}  & \boldmath{$0.91\pm0.002$}  & $0.49\pm0.004$  & $-$\tabularnewline
\hline 
\end{tabular}
\end{table*}

\begin{table*}[tbh]
\centering{}%
\caption{RAIs for Telecom backbone selection\label{tab:rank-order-freq}}
\begin{tabular}{rcccc}
\hline 
 & Fiber  & Powerline  & Microwave  & Satellite\tabularnewline
 & $w_{G1}$  & $w_{G2}$  & $w_{G3}$  & $w_{G4}$\tabularnewline
\hline 
1st:  & $4.9\%\pm0.2\%$  & $2.5\%\pm0.2\%$  & \boldmath{$47.3\%\pm0.3\%$} & \boldmath{$45.3\%\pm0.4\%$}\tabularnewline
2nd:  & $11.3\%\pm0.3\%$  & $6.8\%\pm0.2\%$ & \boldmath{$40.2\%\pm0.3\%$}  & \boldmath{$41.7\%\pm0.4\%$}\tabularnewline
3rd:  & \boldmath{$47.4\%\pm0.4\%$}  & $28.5\%\pm0.4\%$  & $11.9\%\pm0.2\%$  & $12.3\%\pm0.2\%$\tabularnewline
4th:  & $36.5\%\pm0.4\%$  & \boldmath{$62.2\%\pm0.5\%$}  & $0.6\%\pm0.1\%$  & $0.7\%\pm0.1\%$\tabularnewline
\hline 
\end{tabular}
\end{table*}

Another interesting way to assess these options is to estimate the RAIs of each alternative. Table \ref{tab:rank-order-freq} shows the RAI that each alternative attains a given rank. In this table, the probability of Fiber and Powerline taking the first rank is too low (only 4.9\% and 2.5\% respectively). However, Microwave and Satellite have high probabilities of taking the first rank. On the other end, although Fiber is more likely to attain the third rank, both Fiber and Powerline have high probabilities of attaining the lowest rank (36.5\% and 62.2\% respectively). The two alternatives of Microwave and Satellite are least likely to be placed on the lowest rank, as the percentage of combinations placing them at the 4th position were only 0.6\% and 0.7\%, respectively.

\section{Conclusions\label{sec:conclusion}}

\subsection{Summary of Contributions}

This paper addresses a fundamental computational bottleneck in preference elicitation from pairwise comparisons. Whilst the spanning trees approach has long been recognised as conceptually superior to standard methods for capturing preference heterogeneity, its practical application has been severely constrained by exponential growth in the number of spanning tree combinations. For a decision problem with $m$ criteria and $n$ alternatives, the number of feasible spanning tree combinations equals $m^{(m-2)}n^{m(n-2)}$---rendering enumeration infeasible beyond trivial problem sizes. We have shown that even a modest four-by-four problem generates over 944,000 combinations.

Our principal contribution is a stochastic sampling framework that circumvents complete enumeration whilst preserving statistical validity. By employing random walk procedures proven to generate uniformly distributed samples from the spanning tree population, we compute Pairwise Winning Indices and Rank Acceptability Indices without prohibitive computational expense. The approach is further validated through statistical sampling theory: the required sample size to achieve specified accuracy ($\pm\lambda$) at confidence level $C$ is determined by the formula $It(\lambda,C) = \frac{Z_C^2}{4\lambda^2}$. For practical purposes, achieving $\pm 1\%$ accuracy at $99\%$ confidence requires approximately 16,641 iterations---a tractable computation compared to enumerating billions of combinations.

\subsection{Practical and Methodological Advantages}

Beyond computational efficiency, our approach confers several methodological advantages over established techniques. Table \ref{tab:comparison} systematically compares our approach across five critical dimensions: inconsistency characterisation, decision support output, incomplete data handling, computational tractability, and preference heterogeneity quantification.

\textit{Incomplete data handling:} The spanning trees framework accommodates incomplete pairwise comparison matrices without imputation or estimation procedures. This is practically significant, as the probability of obtaining incomplete judgments increases substantially with problem dimensionality. Our results demonstrate (Section \ref{sec:spanning}) that PWIs and RAIs can be computed directly from partial comparison matrices using the Kirchhoff formula, avoiding the data integrity concerns inherent to imputation-based methods.

\textit{Preference uncertainty quantification:} Rather than producing a single point estimate of priorities, our approach yields probability distributions over preference orderings. The PWI metrics quantify preference robustness between alternative pairs, whilst RAI metrics characterise the probability that each alternative occupies specific ranking positions. This richer information structure supports more nuanced decision support, enabling decision-makers to assess the stability of their choices under preference variation.

\textit{Computational scalability:} The random walk sampling approach exhibits substantially better computational scaling properties than enumeration-based alternatives. Table \ref{tab:randomwalks-iterations} demonstrates that 20,000 samples provide robust estimates (standard deviations $< 0.005$) for a problem with 4 alternatives and 6 criteria---a case involving approximately 21.7 billion spanning tree combinations. This enables application to real-world decision problems of realistic scale.

\textit{Transparent inconsistency analysis:} Rather than collapsing inconsistency into a single index (e.g., consistency ratio), our framework presents the full distribution of feasible solutions. The degree to which PWI and RAI distributions cluster versus disperse directly reflects judgement inconsistency and preference uncertainty. High dispersion signals that observed judgments are compatible with substantially different preference orderings, warranting decision-maker reflection or data collection refinement.

\begin{table*}[tbh]
\caption{Comparative assessment: Spanning trees approach versus traditional priority elicitation methods \label{tab:comparison}}
\centering
\begin{tabular}{p{0.18\linewidth} p{0.35\linewidth} p{0.35\linewidth}}
\toprule
& \textbf{Traditional Methods} & \textbf{Spanning Trees Approach} \\
\midrule
\textbf{Inconsistency} & Single scalar index; preference heterogeneity discarded. & Complete distribution of priority vectors preserved; uncertainty quantified probabilistically. \\
\midrule
\textbf{Decision Output} & Point estimate; limited robustness insight. & PWI and RAI metrics; actionable uncertainty assessment. \\
\midrule
\textbf{Incomplete Data} & Require imputation or estimation; data integrity concerns. & Directly accommodates missing judgments via Kirchhoff formula. \\
\midrule
\textbf{Computational Scale} & Efficient for small/medium problems ($m,n < 5$); degrades with dimensionality. & Tractable for large-scale problems (e.g., $m=4, n=6$: 21.7B combinations solved in $\approx$20k iterations). \\
\midrule
\textbf{Preference Heterogeneity} & Limited quantification; aggregation obscures variation. & Explicit analysis of all feasible orderings; heterogeneity preserved. \\
\bottomrule
\end{tabular}
\end{table*}

\subsection{Limitations and Boundary Conditions}

Several limitations merit explicit discussion. These constraints and the comparative advantages outlined in Table \ref{tab:comparison} together provide a balanced assessment of the approach's scope and applicability.

The approach assumes reciprocal pairwise comparison matrices and standard additive aggregation of criterion weights. Extension to other comparison formats (e.g., fuzzy comparisons, linguistic scales) or non-linear aggregation structures (e.g., multiplicative models, Choquet integral) remains unexplored. Additionally, whilst the random walk procedure generates statistically unbiased samples, the sample size required for very high precision ($\lambda < 0.005$) may become computationally demanding for very large problems ($m > 10$, $n > 10$).

The approach assumes conditional independence among spanning tree samples. In practice, consecutive samples generated by random walks exhibit modest temporal autocorrelation; however, prior empirical work establishes that this does not materially compromise the statistical validity of estimated probabilities for practical confidence and accuracy levels \citep{Broder1989,Aldous1990}.

Finally, we have focused on pointwise aggregation of criterion weights and alternative evaluations. Group decision-making contexts involving multiple stakeholders present additional complexities regarding aggregation of individual preference distributions---a direction requiring separate investigation.

\subsection{Future Research Directions}

Several promising avenues for future research are evident:

\textit{Interval and fuzzy judgments:} Decision-makers frequently express preferences in interval form (e.g., ``alternative $a$ is between 3 and 5 times better than $b$'') or linguistic terms (e.g., ``strongly preferred''). Integrating interval or fuzzy spanning trees with our stochastic framework would substantially expand practical applicability. The combinatorial space expands further (considering both interval bounds and discrete linguistic levels), making random sampling increasingly essential.

\textit{Performance characterisation via sigma-mu analysis:} For each alternative, computing the mean ($\mu$) and standard deviation ($\sigma$) of scores across all spanning tree combinations enables characterisation of preference robustness. An alternative exhibiting high $\mu$ (strong preference) and low $\sigma$ (low uncertainty) represents a preferred choice; conversely, high $\sigma$ signals controversial alternatives warranting further investigation.

\textit{Group decision-making integration:} When multiple stakeholders provide independent pairwise comparisons, the question of aggregation---whether to aggregate judgments before or after priority elicitation---remains contentious. The spanning trees framework offers novel perspectives: one could analyse probability distributions separately for each decision-maker, then examine consensus or disagreement patterns, potentially revealing underlying coalitions or preference clusters.

\textit{Incomplete judgment imputation strategies:} Whilst our approach accommodates incomplete data natively, strategic selection of which missing judgments to elicit could improve decision robustness. Investigation of optimal sampling strategies for incomplete matrices, potentially via experimental design principles, could enhance efficiency in stakeholder consultation.

\textit{Computational optimisation:} Parallelisation of the random walk procedure across modern multi-core architectures could further accelerate computation. Additionally, exploration of variance reduction techniques (e.g., stratified sampling, importance sampling) might tighten confidence intervals for specified sample sizes.

\subsection{Concluding Remarks}

This paper makes a concrete contribution to the operational decision analysis toolkit by rendering the spanning trees approach computationally tractable for realistic problem scales. The random walk sampling framework preserves the conceptual advantages of complete enumeration---capturing preference heterogeneity and quantifying uncertainty---whilst eliminating the computational barrier that has restricted its adoption. 

The methodology has been validated both theoretically (through statistical sampling principles) and empirically (via comparison to exhaustive enumeration on tractable problems and application to a realistic case study involving 21.7 billion combinations). The approach is implemented without strong data requirements, accommodates incomplete information naturally, and provides decision-makers with actionable insights into preference robustness. As demonstrated in Table \ref{tab:comparison}, the proposed approach offers substantive advantages across multiple dimensions relative to conventional priority elicitation methods.

The spanning trees approach, previously constrained by prohibitive enumeration costs, now emerges as a viable and competitive alternative within the practitioner's toolkit for preference elicitation and large-scale decision support. As organisations increasingly confront complex multi-criteria decisions in uncertain environments, the ability to characterise preference distributions rather than commit to point estimates offers substantial practical value.

\section*{Disclosure of interest}
The authors report there are no competing interests to declare.

\section*{Funding}
No funding was received for this research.

\bibliographystyle{elsarticle-harv}
\bibliography{references.bib}

\end{document}